\documentclass{elsarticle}
\usepackage[english]{babel}
\usepackage{amsmath,amsfonts,amssymb}
\usepackage{listing}

\begin{document}
\title{Modelling Gamma-Ray Photon Emission \& Pair Production in High-Intensity Laser-Matter Interactions}
\renewcommand{\thefootnote}{\fnsymbol{footnote}}
\author{C.P. Ridgers$^{1,2}$\footnote[2]{Present address: Department of Physics, The University of York, Heslington, York, YO10 5DD, UK}, J.G. Kirk$^3$, R. Duclous$^4$, T. Blackburn$^{1}$, C.S. Brady$^5$, K. Bennett$^5$, T.D.Arber$^5$, A.R. Bell$^{1,2}$}
\address{$^1$Clarendon Laboratory, University of Oxford, Parks Road, Oxford, OX1 3PU, UK \\
$^2$Central Laser Facility, STFC Rutherford-Appleton Laboratory, Chilton, Didcot, Oxfordshire, OX11 0QX, UK \\
$^3$Max-Planck-Institut f\"{u}r Kernphysik, Postfach 10 39 80, 69029 Heidelberg, Germany
$^4$Commissariat \`{a} l'Energie Atomique, DAM DIF, F-91297 Arpajon, France  \\
$^5$Centre for Fusion, Space and Astrophysics, University of Warwick, Coventry, CV4 7AL, UK \\}

\begin{abstract}
In high-intensity ($>10^{21}$Wcm$^{-2}$) laser-matter interactions gamma-ray 
photon emission by the electrons can strongly effect the electron's dynamics 
and copious numbers of electron-positron pairs can be produced by the emitted 
photons.  We show how these processes can be included in simulations
by coupling a Monte-Carlo algorithm describing the emission to a 
particle-in-cell code.  The 
Monte-Carlo algorithm includes quantum corrections to the photon emission, 
which we show must be included if the pair production rate is to be
correctly determined.  The accuracy, convergence and energy conservation 
properties of the Monte-Carlo algorithm are analysed in simple test problems.  
\end{abstract}

\maketitle

\section{Introduction}

\setcounter{footnote}{0}

High power lasers, operating at intensities $I>10^{21}$Wcm$^{-2}$, create 
extremely strong electromagnetic fields ($E_L\gtrsim10^{14}$Vm$^{-1}$).  These 
fields can accelerate electrons sufficiently violently that they radiate a 
large fraction of their energy as gamma-rays within a single 
laser cycle.  As a result the radiation reaction force becomes important in 
determining the electron trajectories \cite{Dirac_38}.  In addition, quantum 
aspects of the radiation emission are important 
\cite{Kirk_09,Sokolov_10,Duclous_11,Elkina_11} and the emitted photons readily 
produce 
electron-positron pairs \cite{Bell_08}.  Gamma-ray photon and pair production 
can be investigated with today's petawatt-power lasers in specially arranged 
experiments.  Furthermore, these emission processes will dominate the 
dynamics of plasmas generated by next generation 10PW lasers 
\cite{Nerush_11,Sokolov_11,Ridgers_12}.  In 10PW laser-plasma interactions 
the QED emission processes and the plasma physics 
processes are strongly coupled.  The resulting plasma is best defined as a 
`QED-plasma', partially analogous to those thought to exist in extreme 
astrophysical environments such as the magnetospheres of pulsars \& active 
black holes \cite{Goldreich_69}. 
It is therefore highly desirable that gamma-ray photon emission and pair 
production be included in laser-plasma simulation codes.  In this paper we 
will describe how these processes may be simulated using a Monte-Carlo 
algorithm \cite{Duclous_11} and how this algorithm can be coupled to a 
particle-in-cell (PIC) code \cite{Dawson_62}, allowing self-consistent 
simulations of QED-plasmas.  

Several PIC codes have been modified to include a classical description of 
gamma-ray emission and the resulting radiation reaction \cite{Zhidkov_02}. 
The neglect of quantum effects limits the range of validity of such codes.  
The parameter which determines the importance of quantum effects in emission 
by an electron is $\eta=E_{RF}/E_s$ where $E_{RF}$ is the electric field in the
electron's rest frame and $E_s=1.3\times10^{18}$Vm$^{-1}$ is the Schwinger field 
required to break down the vacuum into electron-positron pairs 
\cite{Sauter_31}.  When $\eta\sim1$: (i) classical theory predicts
 unphysical features, such as the emission of photons with more energy than 
the parent electron.  Quantum modifications to the radiated spectrum are, 
therefore, essential \cite{Erber_66,Sokolov_68}.  (ii) A quantum description 
of photon 
emission is probabilistic and as a result the electron motion becomes 
stochastic \cite{Shen_72}.  (iii) The emitted photons are sufficiently 
energetic to readily produce electron-positron pairs \cite{Erber_66}.  
These pairs go on to generate photons and thus further pairs, initiating 
a cascade of pair production \cite{Bell_08}.  

The importance of quantum effects in current and next-generation laser-matter 
interactions can be estimated by assuming that $E_{RF}\sim \gamma E_L$, 
where $E_L$ is the 
laser's electric field and $\gamma$ is the Lorentz factor of the electrons 
in the laser fields.  For current 1PW lasers (intensity 
$I\sim10^{21}$Wcm$^{-2}$) $E_L/E_s\sim10^{-4}$.  To reach 
$\eta>0.1$, $\gamma>1000$ is required.  The laser pulse 
typically accelerates electrons to $\gamma\sim{}a$, where 
$a=eE_L\lambda_L/2\pi m_e c^2$ is the strength parameter of the laser wave 
($\lambda_L$ is the laser wavelength).  
For $I=10^{21}$Wcm$^{-2}$, $a=30$ and so in order to observe quantum effects 
the electrons must be accelerated to 
high energies externally.  GeV electron beams, which can now be generated by 
laser-wakefield acceleration \cite{Leemans_06}, are sufficient.  The 
collision of such a beam with a 1PW laser pulse could reach the $\eta>0.1$ 
regime \cite{Sokolov_10_2,Thomas_12}.  In fact this regime has recently been 
reached in similar experiments where an energetic electron beam, produced by 
a particle accelerator, interacts with strong crystalline fields 
\cite{Andersen_12}\footnote{Other experiments have been performed where: a 
particle accelerator produced a beam of electrons with 46.6GeV which 
subsequently collided with a laser pulse of $I=10^{18}-10^{19}$Wcm$^{-2}$ 
\cite{Bula_96}; (2) laser wakefield acceleration produced 100MeV electrons 
which were collided with a pulse of $I=5\times10^{18}$Wcm$^{-2}$ 
\cite{TaPhouc_12}.  However due to the low laser intensity the radiation 
(and pair production in the former experiment) are in a substantially 
different regime to those considered here.}.  

In the case where the electron beam is externally accelerated and 
then collided with a laser pulse, the plasma processes which cause the 
acceleration and the gamma-ray \& pair emission during the collision are 
decoupled 
and may be considered separately.  The same is true of recent laser-solid 
experiments where photon and pair production occur in the electric fields of 
the nuclei of high-Z materials far from the laser focus \cite{H_Chen_10}.  
By contrast, in laser-solid interactions at intensities expected to be reached 
by next-generation 10PW lasers ($>10^{23}$Wcm$^{-2}$ \cite{Mourou_07}) 
$E_L/E_s\gtrsim10^{-3}$ and $a\gtrsim100$ and so the laser pulse itself can 
accelerate electrons to high enough energies to reach 
$\eta>0.1$.  In this case the 
emission \& plasma processes both occur in the plasma generated at the laser 
focus.  The rates of the QED emission processes for a given electron in
the plasma depend on the local electromagnetic fields and the electron's 
energy, which are determined by the plasma physics processes.  Conversely, 
the QED emission processes can alter the plasma currents and so affect the 
plasma physics. As a result the macroscopic plasma processes and the QED 
emission processes cannot be considered separately in the resulting QED-plasma.

In this paper we will describe a Monte-Carlo algorithm for calculating the 
emission of gamma-ray photons and pairs in strong laser fields.  In addition 
to being more widely applicable than a classical description of the emission, 
this quantum description of emission in terms of discrete particles is more 
suited to coupling to a Particle-in-Cell code.  We will detail how this 
coupling can be achieved so as to self-consistently model the feedback between 
the plasma and emission processes; we will refer to the coupled code as 
`QED-PIC' for brevity.  QED-PIC codes based on this or a similar technique 
have recently been employed for simulations of both laser-plasma interactions 
\cite{Nerush_11,Sokolov_11,Ridgers_12} and pulsar magnetospheres 
\cite{Timokhin_10}.   

\section{The Emission Model}
\label{emission_section}

\begin{figure}
\centering
\includegraphics[scale=0.5]{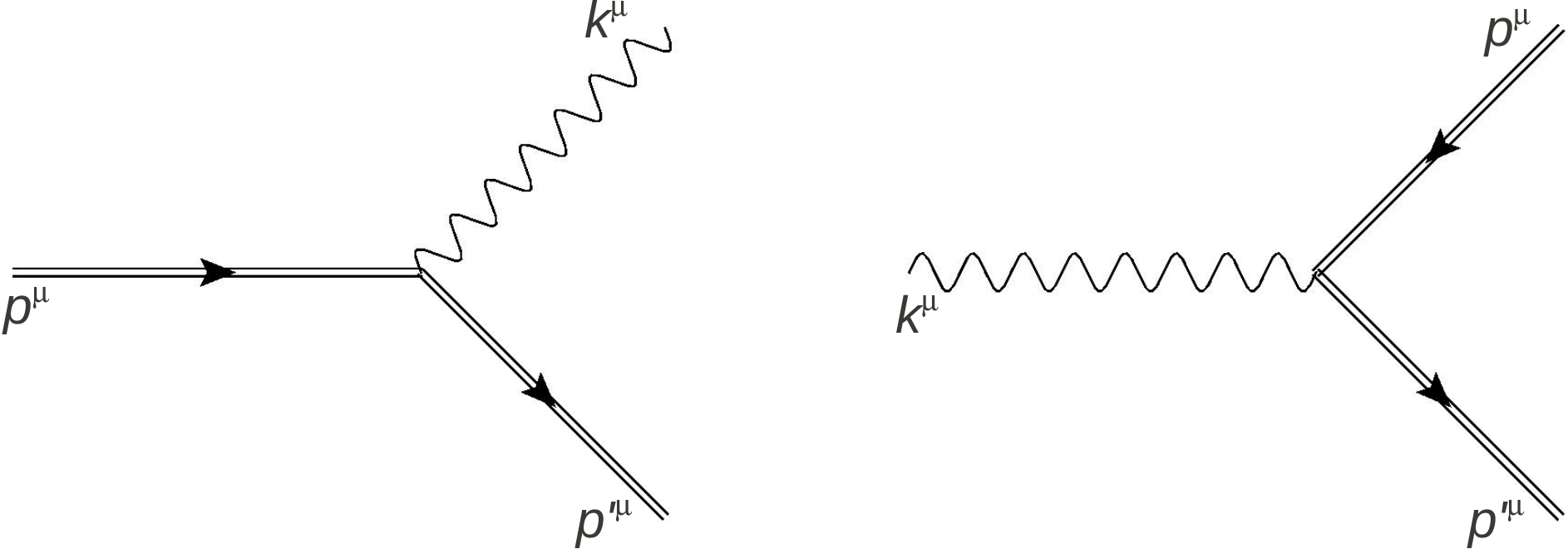}
\caption{\label{Feyn_diags} Diagrammatic representation of the emission 
processes included in the model: photon emission by an electron (left) \& pair 
production by a photon (right).  The double lines represent `dressed' states.}
\end{figure}

The emission model described here is detailed in Refs. 
\cite{Kirk_09} \& \cite{Duclous_11}.  For completeness, we will 
summarise the important details in this section.  
The electromagnetic field is split into high \& low frequency components.  The
low frequency macroscopic fields (the `laser fields') vary on scales similar 
to the laser 
wavelength and are coherent states that are unchanged in QED interactions. 
These fields behave classically \cite{Glauber_63} and are computed by solving 
Maxwell's equations including the plasma charges and currents smoothed on this 
length scale.  Interactions between electrons, positrons and the high-frequency 
component of the electromagnetic field (gamma-rays) can be included using the 
method described by Baier and Katkov \cite{Baier_68}, in which 
particles (electrons, positrons and photons) move classically in between 
point-like QED interactions. The interaction probabilities are calculated 
using the strong-field or Furry representation \cite{Furry_51}, in which the 
charged particle basis states are \lq dressed\rq\ by the laser fields.  
Feynman diagrams for the dominant 
first-order (in the fine-structure constant $\alpha_f$) interactions included 
in the model are 
shown in figure \ref{Feyn_diags} and represent: the emission of a gamma-ray 
photon by an electron accelerated by the laser fields (the equivalent process 
of photon emission by a positron 
is also included in the model) \& the creation of an electron-positron pair 
by a gamma-ray photon interacting with the laser fields.

This approach rests on two approximations:

1. The macroscopic laser fields are treated as static during the QED 
interactions, i.e., \lq instantaneous\rq\ and \lq local\rq\ values of the 
transitions rates are calculated. This approximation holds if the coherence 
length\footnote{In the classical picture the coherence length is the 
path length over which an electron (of energy $\gamma m_e c^2$.) is deflected 
by $1/\gamma$} associated with the interaction is small compared to 
$\lambda_{\rm L}$. In the case of a monochromatic plane wave, the coherence 
length is $\lambda_{\rm L}/a$ [15], so the approximation is valid for $a\gg1$.

2. The laser fields are much weaker than the Schwinger field.  In this case 
we may make the approximation that the emission rates depend only on the 
Lorentz-invariant parameters: $\eta=(e\hbar/m_e^3c^4)|F_{\mu\nu}p^{\nu}|=E_{RF}/E_s$ 
and $\chi=(e\hbar^22m_e^3c^4)|F^{\mu\nu}k_{\nu}|$, $p^{\mu}$ ($k^{\mu}$) is the 
electron's (photon's) 4-momentum; and are independent of the Lorentz invariants 
$\mathcal{F}=|E^2-c^2B^2|/E_s^2$ \& 
$\mathcal{G}=|\mathbf{E}\cdot c\mathbf{B}|/E_s^2$  associated with the laser 
fields.  This requires
$\eta^2,\chi^2\gg\mbox{Max}[\mathcal{F},\mathcal{G}]$ \& 
$\mathcal{F},\mathcal{G}\ll1$.  For next 
generation 10PW laser pulses $\eta,\chi \sim O(1)$ \& $E_L/E_s\sim10^{-3}$ and 
so this approximation
does not unduly limit the validity of the model.  Under this `weak-field' 
approximation the emission rates in the particular macroscopic field 
configuration
approximately equal those in any other configuration with 
$\mathcal{F},\mathcal{G}\ll 1$, provided the configurations share the same 
values of $\eta$ and $\chi$. Furthermore, this approximation ensures that the 
particle dynamics in between QED interactions may be treated classically 
\cite{Baier_68}. Convenient configurations are a uniform, static magnetic field
\cite{Erber_66} and a plane wave \cite{Ritus_85}.  The reaction rates for 
the processes shown in figure \ref{Feyn_diags} in these field configurations 
are well-known.  Here we use the nomenclature of the the static magnetic field 
case in which: photon emission corresponds to synchrotron radiation, also 
called magnetic bremsstrahlung and pair creation corresponds to magnetic pair 
production\footnote{In the plane-wave case these processes correspond to 
nonlinear Compton scattering and multiphoton Breit-Wheeler pair production.}. 

\subsection{Synchrotron Radiation}
\label{radiation_section}

The (spin \& polarisation averaged) rate of emission of gamma-ray photons by an
electron (or positron) in a constant magnetic field, where $cB \ll E_s$, is 

\begin{equation}
\label{rate_photons_1}
\frac{d^2 N_{\gamma}}{d\chi dt} = \frac{\sqrt{3}\alpha_fc}{\lambda_c}\frac{cB}{E_s}\frac{F(\eta,\chi)}{\chi}
\end{equation}

The electron's energy is parameterised by $\eta$ and 
the photon's energy 
by $\chi$.  $\lambda_c$ is the Compton wavelength \& $F(\eta,\chi)$ is the 
quantum-corrected synchrotron spectrum as given by Erber \cite{Erber_66} 
and Sokolov \& Ternov \cite{Sokolov_68}.  $F(\eta,\chi)$ is
reproduced in
\ref{photon_emissivity_section}.  The 
modification of $F(\eta,\chi)$ away from the classical synchrotron spectrum 
leads to a quantum 
correction to the instantaneous power radiated 
$\mathcal{P}=(4\pi{}m_ec^3/3\lambda_c)\alpha_f\eta^2g(\eta)=\mathcal{P}_cg(\eta)$.  $\mathcal{P}_c$ is the classical power and 
$g(\eta)\approx [1+4.8(1+\eta)\ln(1+1.7\eta)+2.44\eta^2)]^{-2/3}$.  Note that when $\eta=1$, $g(\eta)=0.2$ 
and $\mathcal{P}$ is reduced by a factor of five.    

Equation (\ref{rate_photons_1}) can be written in terms of $\eta$ by 
multiplying through by $\gamma$ and identifying $\eta=\gamma cB/E_s$.  
Integrating over $\chi$ yields

\begin{equation}
\label{rate_photons}
\frac{dN_{\gamma}}{dt} = \frac{\sqrt{3}\alpha_fc}{\lambda_c}\frac{\eta}{\gamma}h(\eta) = \lambda_{\gamma}(\eta)
\end{equation}

 The exact forms for $g(\eta)$ \& $h(\eta)$ are given in 
\ref{photon_emissivity_section}.  The alternative, more 
compact, symbol $\lambda_{\gamma}$ will be more convenient to use in the 
later equations (\ref{master_electrons}) \& (\ref{master_photons}).  
The probability that a photon is emitted with a 
given $\chi$ (by an electron with a given $\eta$) is 
$p_{\chi}(\eta,\chi)=[1/h(\eta)][F(\eta,\chi)/\chi]$.  The emitting electron or
positron experiences a recoil in the emission which, ignoring momentum
transferred to the background field, balances the momentum of the emitted 
photon.

\subsection{Magnetic Pair Production}
\label{pair_prod}

The probability of photon absorption through magnetic pair creation (averaged 
over spin and polarisation) in a constant magnetic field can be written 
in terms of a differential optical depth \cite{Erber_66}

\begin{equation}
\label{pair_prod_rate}
\frac{d\tau_{\pm}}{dt} = \frac{2\pi\alpha_fc}{\lambda_c}\frac{m_ec^2}{h\nu_{\gamma}}\chi{}T_{\pm}(\chi) = \lambda_{\pm}(\chi)
\end{equation}

$h\nu_{\gamma}$ is the energy of the gamma-ray photon generating the 
electron-positron pair and $\chi$ is the corresponding value of this parameter.
  $T_{\pm}(\chi)$ controls the pair emissivity and is given in 
\ref{pair_emissivity_section}. 

The energy of the photon is split between the generated 
electron \& positron (again ignoring momentum transferred to the macroscopic 
field). 
The probability that 
one member of the emitted pair has a fraction $f$ of the photon's energy 
(parameterised by $\chi$) is
$p_f(f,\chi)$ \cite{Daugherty_83}.  This function is also given in 
\ref{pair_emissivity_section}.

\subsection{Quasi-Classical Kinetic Equations}
\label{master_equations}

As mentioned above, the weak-field assumption ensures that the motion of the 
electron between emission events can be treated classically. However, the 
emission itself causes a quantum effect 
on the motion known as `straggling' 
\cite{Duclous_11,Shen_72}.  In the quantum 
description emission is probabilistic rather than deterministic as in the 
classical picture.  Considering photon emission, this leads to a stochastic 
recoil of the emitting electron or positron, which gives rise to a quantum 
equivalent of 
the classical radiation reaction force \cite{DiPiazza_10}.  The stochastic 
nature of this reaction force allows some electrons \& positrons 
(the `stragglers') to access classically inaccessible regions of phase-space.   

The straggling effect may be quantified as follows.  If 
$f_{\pm}(\mathbf{x},\mathbf{p},t)d^3\mathbf{x}d^3\mathbf{p}$ is the probability 
that an electron or positron is at phase space coordinates 
$(\mathbf{x},\mathbf{p})$ and 
$f_{\gamma}(\mathbf{x},\mathbf{k},t)d^3\mathbf{x}d^3\mathbf{k}$ is the 
equivalent for a photon, then these distribution functions obey the following 
quasi-classical kinetic equations \cite{Elkina_11,Shen_72,Sokolov_10_2}.

\begin{align}
\label{Vlasov_ep}
\frac{\partial{}f_{\pm}(\mathbf{x},\mathbf{p},t)}{\partial{}t} + \mathbf{v}\cdot\nabla f_{\pm}(\mathbf{x},\mathbf{p},t) + \mathbf{F}_L\cdot\nabla_{\mathbf{p}}f_{\pm}(\mathbf{x},\mathbf{p},t) = \left(\frac{\partial{}f_{\pm}}{\partial{}t}\right)_{em} \\
\label{Vlasov_gamma}
\frac{\partial{}f_{\gamma}(\mathbf{x},\mathbf{k},t)}{\partial{}t} + c\hat{\mathbf{v}}\cdot\nabla f_{\gamma}(\mathbf{x},\mathbf{k},t)= \left(\frac{\partial{}f_{\gamma}}{\partial t}\right)_{em}
\end{align}  

The left-hand side of equation (\ref{Vlasov_ep}) describes the classical 
propagation of electrons \& positrons in the macroscopic fields between 
emissions.  The left-hand side of equation (\ref{Vlasov_gamma}) describes 
photons following null geodesics, i.e. propagating at $c$ in direction 
$\hat{\mathbf{v}}$.  In each equation the right-hand side describes the 
emission processes.  
As stated above, for $a \gg 1$ the emission
 occurs on very small scales compared to the variations in the macroscopic 
field and so is effectively point-like, depending only on the local fields.  
Photon absorption and electron-positron annihilation have been ignored.

 We assume that for a system initially containing $N$ electrons, the $N$ 
particle distribution function $f^N_{-}$ can be assumed to be equal to $Nf_{-}$.
This holds as the peak in the emitted photon energy spectrum is at much higher
energy than the energy of the laser photons.  
As a result, the motion of the emitting electrons is not 
correlated on the length scales equal to the wavelength of the gamma-ray 
photons (except at the very lowest energies, where we assume a negligible 
amount of radiation is emitted) and the emission is incoherent.

\subsection{Energy Spectra in Simple Macroscopic Field Configurations}

In section \ref{Monte_Carlo} we will discuss how equations (\ref{Vlasov_ep}) \&
(\ref{Vlasov_gamma}) may be solved 
numerically in a general electromagnetic field configuration using a 
Monte-Carlo algorithm.  However, insight may be gained by considering the 
following specific field configurations, which will provide test problems 
for the  Monte-Carlo algorithm: a constant and homogeneous magnetic field and 
a circularly polarised electromagnetic wave. 

For propagation perpendicular to a uniform \& static magnetic field, 
the controlling parameters $\eta$ \& $\chi$ depend only on the emitting 
particle's energy and the strength of the magnetic field $B$. We define the 
energy distribution as 
$\Phi_{\pm,\gamma}(\gamma,t)=m_ec\int d^3\mathbf{x}d^2\Omega{}f_{\pm,\gamma}(\mathbf{x},\mathbf{p},t)p^2$; 
where $d^2\Omega$ is the element of solid angle in momentum space, 
$\Phi_{-}(\gamma,t)d\gamma$ is the probability that an electron has Lorentz 
factor $\gamma$ \& $\Phi_{\gamma,}(\epsilon,t)d\epsilon$ is the probability that 
the photon has energy $\epsilon=h\nu_{\gamma}/m_ec^2$.  The equations for the 
evolution of these distribution functions are \cite{Shen_72,Sokolov_10_2}  

\begin{align}
\label{master_electrons}
\frac{\partial{}\Phi_{\pm}(\gamma,t)}{\partial{}t}=-\lambda_{\gamma}(\eta)\Phi_{\pm}(\gamma,t) +& \int_{\gamma}^{\infty}d\gamma'\lambda_{\gamma}(\eta')p_{\chi}(\eta',\chi)\Phi_{\pm}(\gamma',t) \nonumber \\
&+ \int_{\chi_1}^{\infty}d\chi\lambda_{\pm}(\chi)p_f(f,\chi)\Phi_{\gamma}(\epsilon,t) \\
\label{master_photons}
\frac{\partial{}\Phi_{\gamma}(\epsilon,t)}{\partial{}t} = -\lambda_{\pm}(\chi)\Phi_{\gamma}(\epsilon,t)& + \int_{\gamma_1}^{\infty}d\gamma{}\lambda_{\gamma}(\gamma)(\eta)p_{\chi}(\eta,\chi)[\Phi_{-}(\gamma,t)+\Phi_{+}(\gamma,t)]
\end{align}

$p_{\chi}$ is the probability that an electron or positron with parameter 
$\eta$ emits a photon with $\chi=(\gamma'-\gamma)(cB/2E_s)$; 
$p_f$ is the probability that on pair creation, the photon gives a fraction 
$f=\eta/2\chi$ of its energy to the positron \cite{Daugherty_83}.  
The lower limits of the integrals $\gamma_1=(2\chi)/(cB/E_s)$ \& 
$\chi_1=(\gamma/2)(cB/E_s)$ arise from energy conservation; an electron 
(photon) cannot emit a photon (electron-positron pair) with more energy than 
it possesses.  It should be noted that equations (\ref{master_electrons}) \& 
(\ref{master_photons}) are only valid for ultra-relativistic electrons \& 
positrons emitting synchrotron-like radiation; therefore $\Phi_{\pm}$ \& 
$\Phi_{\gamma}$ become unreliable below some energy, where the radiation will 
not be synchrotron-like but can also be assumed to be unimportant.  

For the case of an electron with initial $\gamma_0$ counter-propagating 
relative to a plane circularly-polarised electromagnetic wave of strength 
parameter $a$, where $\gamma_0\gg a$, then the energy gained by the electron 
from the wave may be ignored and the distribution functions are also 
described by equations (\ref{master_electrons}) \& (\ref{master_photons}).  In 
this case $\eta=2\gamma{}E/E_s$ \& $\chi=(h\nu_{\gamma}/m_ec^2)(E/E_s)$, where
 $E$ is the wave's electric field.  

Multiplying equation (\ref{master_electrons}) by $\gamma$ and integrating over
$\gamma$ (neglecting the 
contribution from pair production)  
yields

\begin{equation}
\label{Losses_QED}
\frac{d\langle\gamma\rangle}{dt} = -\left\langle\frac{\mathcal{P}(\eta)}{m_ec^2}\right\rangle
\end{equation}    

For comparison the equation of motion for a particle radiating in a 
deterministic fashion is

\begin{equation}
\label{Losses_classical}
\frac{d\gamma_d(t)}{dt} = -\frac{\mathcal{P}[\eta_d(t)]}{m_ec^2}
\end{equation}        

$\gamma_d(t)$ \& $\eta_d(t)$ are the Lorentz factor 
and $\eta$-parameter of the electron moving on a deterministic worldline.  
To arrive at this equation we have followed the Landau \& Lifshitz 
prescription for dealing with the radiation reaction force \cite{Landau_87} 
and taken the ultra-relativistic limit.  We have also made the substitution 
$\mathcal{P}_c\rightarrow\mathcal{P}$ \cite{Kirk_09}, thus capturing the 
quantum reduction in the synchrotron power but not the stochasticity of the 
emission. Henceforth this will be described as the `deterministic' emission 
model, as opposed to the `probabilistic' model which includes the quantum 
stochasticity.  In the classical limit the variance in $\Phi_{-}$ is small, 
$\Phi_{-}\rightarrow\delta[\gamma-\gamma_d(t)]$, 
$d\langle{}\gamma\rangle/dt\rightarrow d\gamma_d/dt$ and 
$\langle{}\mathcal{P}(\eta)\rangle\rightarrow\mathcal{P}[\eta_d(t)]$, 
demonstrating correspondence between the probabilistic equation 
(\ref{Losses_QED}) and the deterministic equation (\ref{Losses_classical}).

\subsection{Feedback on the Macroscopic Fields}
\label{feedback}

So far we have discussed emission in constant classical fields.  However, one 
of the defining features of a QED-plasma is that the emission processes can 
influence these fields.  Radiation reaction exerts a drag force, altering 
the velocity of the electrons and positrons and so altering the current in the 
plasma.  Although no net current is produced in the pair production process, 
it acts as a source of current carriers.  Subsequent 
acceleration by the background fields separates the electron and positron 
which then alter the current in the plasma.  
These modifications to the current effect the evolution of the macroscopic 
fields.

\section{The Monte-Carlo Algorithm}
\label{Monte_Carlo}

In this section the Monte-Carlo emission algorithm introduced in Ref. 
\cite{Duclous_11} will be summarised.  This algorithm solves equations 
(\ref{Vlasov_ep}) \& (\ref{Vlasov_gamma}) for the 
distribution functions $f_{\pm}$ and $f_{\gamma}$, capturing the probabilistic
nature of the emission.  The cumulative probability of emission after a 
particle traverses a plasma of optical depth $\tau_{em}$ is 
$P(t)=1-e^{-\tau_{em}}$.  Each macroparticle is assigned an optical depth at 
which it emits by the following procedure.  First $P$ is assigned a 
pseudo-random value between 0 and 1.  The equation for $P$ above is then 
inverted to yield $\tau_{em}$.  For each particle the optical depth evolves
according to 
$\tau(t)=\int_0^t\lambda[\eta(t')]dt'$; $\lambda$ is the appropriate rate of 
emission.  This equation is 
solved numerically by first-order Eulerian integration, i.e. 
$\tau(t+\Delta{}t) = \tau(t) + \lambda(t)\Delta{}t$.  As the macroscopic 
fields are quasi-static, the emission is assumed 
point-like and the rates $\lambda$ depend on the local values of the 
electromagnetic fields and the particle's energy; these are provided by the 
PIC code.  The values of the functions $h(\eta)$ \& $T_{\pm}(\chi)$ are found 
by linearly interpolating the values stored in look-up tables.  When the 
condition $\tau=\tau_{em}$ is met the particle emits.
 
The energy of an emitted photon is 
obtained from the cumulative probability 
$P_{\chi}(\eta,\chi)=\int_0^{\chi}d\chi'p_{\chi'}(\eta,\chi')$, which is tabulated.
  $P_{\chi}$ is assigned to the emitted photon pseudo-randomly in the range 
[0,1].  The value of $\chi$ to which this corresponds at the emitting 
particle's value of $\eta$ is linearly interpolated from tabulated values of
 $P_{\chi}$.  The table 
is cut off at a minimum photon energy, chosen such that the energy of the 
ignored photons sums to no more than $10^{-9}$ times the energy summed over the 
spectrum at the corresponding value of $\eta$.  Photons below the cut-off can 
therefore be safely ignored.

The emitted photon is added to the simulation and assigned an optical depth at 
which it will create a 
pair.  The emitting electron or positron 
recoils, the final momentum $\mathbf{p}_f$ being calculated by subtracting the
photon's momentum from the initial momentum $\mathbf{p}_i$: 
$\mathbf{p}_f=\mathbf{p}_i-(h\nu_{\gamma}/c)\hat{\mathbf{p}}_i$.  When 
the photon's optical depth reaches the assigned value  it creates a pair and 
is annihilated.  Its energy must be shared between the electron and positron 
in the pair.  The cumulative probability that the positron has energy 
$fh\nu_{\gamma}$,
$P_f(f,\chi)=\int_0^{f}df'p_{f'}(f',\chi)$, is tabulated as a function of $f$ 
and $\chi$; $f$ is selected by the same procedure as the photon energy in 
gamma-ray emission.  The pair is then added to the simulation.

The look-up tables used by the Monte-Carlo algorithm for $h(\eta)$, 
$T_{\pm}(\chi)$, $P_{\chi}(\eta,\chi)$ \& $P_f(f,\chi)$ are provided as 
supplementary online data.

\subsection{Accuracy, Numerical Convergence \& Energy Conservation}
\label{numerical_constraints}

Here we discuss the time-step constraints, convergence \& conservation 
properties of the Monte-Carlo algorithm.  In the Monte-Carlo simulation each 
particle can only emit once in a time-step $\Delta{}t$.  There is a finite 
probability, however, that a particle should emit multiple times over the 
duration $\Delta{}t$.  
To minimise this we require $\Delta{t}/\Delta{t_{QED}}\ll1$, where: 
$\Delta{t_{QED}}=1/\mbox{Max}[\lambda_{\gamma}]$ and 
Max$[\lambda_{\gamma}]=(2\sqrt{3}\alpha_fc/\lambda_c)[\mbox{Max}(E,cB)/E_s]
h_0$ is the maximum possible rate of photon emission; 
$\mbox{Max}(E,cB)$ is the maximum electromagnetic field and 
$h_0=5.24$.  
In a situation where prolific pair production occurs one might conclude 
that the time-step constraint $\Delta{}t\lambda_{\pm}\ll1$ is also important.  
However, this constraint is always an order of magnitude less stringent than 
that for photon production\footnote{$\mbox{Max}(\lambda_{\pm})=(2\pi\alpha_fc/\lambda_c)[\mbox{Max}(E,cB)/
E_s]\mbox{Max}(T_{\pm})$.  Here $\mbox{Max}(T_{\pm})=0.2$ and so 
$\mbox{Max}(\lambda_{\gamma})/\mbox{Max}(\lambda_{\pm})=(\sqrt{3}/\pi)
[h_0/\mbox{Max}(T_{\pm})]\sim10$ and photon emission always sets the 
time-step constraint.}.

Sufficient macroparticles must be used that 
$\Phi_{-}$, $\Phi_{\gamma}$ \& $\Phi_{+}$ are adequately sampled.  The number of 
particles $N$ required depends on the specific physical quantity which is 
being examined.  Quantities such as the total energy emitted as gamma-ray 
photons and pairs vary between simulations with a 
standard deviation $\sigma_{N}=\sigma/\sqrt{N}$, where $\sigma$ is the 
standard deviation of $\Phi_{\gamma}$ \& $\Phi_{+}$.  To accurately estimate the 
total energy $E$ we must use sufficient macroparticles that 
$\sigma_{N}/E\ll1$.  If one is interested in details of the spectra 
$\Phi_{\pm}$ \& $\Phi_{\gamma}$, then more macroparticles must be used.  
   
Many more particles than originally present may be generated over the course 
of the simulation.  This may be addressed by: deleting photons immediately 
after generation; merging macroparticles when the number becomes too large.  
The former is the approach we adopt when pair production is negligible; the 
latter is not implemented and is important in the simulation of 
electron-positron cascades as discussed in Ref. \cite{Nerush_11}. 

While the numerical scheme conserves momentum it does not exactly 
conserve energy.  The fractional error in energy conservation during photon 
emission is 
$\Delta\gamma/\gamma_i\approx(1/2\gamma_i)(1/\gamma_f-1/\gamma_i)$; for 
$\gamma_i,\gamma_f \gg 1$.  Here 
 $\gamma_i$ \& $\gamma_f$ are the initial \& final Lorentz factor of the 
electron or positron \cite{Duclous_11}.  The equivalent result for pair 
creation is 
$\Delta\epsilon/\epsilon_{\gamma}\approx(1/2\epsilon_{\gamma})(1/\gamma_-+1/\gamma_+)$ 
when a $\gamma$-ray photon of energy $\epsilon_{\gamma}m_ec^2$ generates an electron 
of Lorentz factor $\gamma_-\gg1$ and a positron of Lorentz factor 
$\gamma_+\gg1$.  These errors in energy conservation are negligibly small 
for $\gamma_i\gg1$ and $\epsilon_{\gamma}\gg1$, which are satisfied in practically
 all
 emission events and therefore the fractional error summed over a large number 
of events is also small.  The errors arise because in reality a small amount of 
momentum is transferred to the classical fields.

\subsection{Coupling the Emission Algorithm to a PIC Code: QED-PIC}

\label{QED_PIC}

The Monte-Carlo emission algorithm can be used to simulate the laser-electron 
beam collision experiments described in the introduction.  It is however 
necessary to include the feedback on the classical macroscopic fields, 
described in section \ref{feedback}, when simulating QED-plasmas generated 
by higher intensity laser pulses.  This can be done by coupling the 
Monte-Carlo emission algorithm to a PIC code.  The basis of the PIC 
technique \cite{Dawson_62}, i.e. the representation of the plasma as 
macroparticles (each representing many real particles), is well suited to 
coupling to the Monte-Carlo code.  
Feedback between particle motion and the electromagnetic fields is calculated 
self-consistently by: interpolating the charge and current 
densities resulting from the positions and velocities of the macroparticles 
onto a spatial grid (this depends on the particle's `shape function'); solving 
Maxwell's equations for the $\mathbf{E}$ \& $\mathbf{B}$ fields; and then 
interpolating these fields onto the particle's positions and pushing the 
particles using the Lorentz force law.  PIC therefore already includes the 
terms on the left-hand side of equation (\ref{Vlasov_ep}) for the electrons 
\& positrons during the particle-push as well as the classical evolution of 
the macroscopic 
fields.  The emission is included by including the Monte-Carlo algorithm in 
the PIC as a new step at the end of each time-step. During emission 
macrophotons and macropairs are produced which represent the same number of 
real photons, electrons or positrons as the emitting macroparticle and 
have the same shape function.  

The modification to the particles velocity caused by radiation reaction is
carried over to the next time-step, in which the macroscopic electromagnetic 
fields then separate the pairs.  The effect of both radiation 
reaction \& pair production on the macroscopic plasma currents is therefore
 included 
when Maxwell's equations are solved in the next time-step, ensuring that the 
interplay of plasma physics effects and the QED emission is simulated 
self-consistently.

The additional time-step constraint $\Delta{}t_{QED}$ introduced by the 
Monte-Carlo algorithm can be compared to those already present in the PIC code.
The Courant-Friedrichs-Lewy condition that information cannot propagate across 
more than one grid-cell (size $\Delta x$) in a single time-step must be 
satisfied.  As the maximum propagation speed is $c$, this gives 
$c\Delta{}t_{CFL}=\Delta{}x=\lambda_L/n$ if $n$ cells are used to 
resolve the laser wavelength.  We require 
$\Delta t <\Delta t_{CFL}$.  The Debye length $\lambda_D$ of the plasma must be 
resolved\footnote{Although this condition can be relaxed for high-order 
shape functions, the time step will still be limited to a multiple of 
$\lambda_D/c$.}.  In which case $\Delta t<\lambda_D/c$.  Assuming 
$\mbox{Max}[E,cB]=E_L$, where $E_L$ is the electric field of the laser, yields 

\begin{equation}
\label{ratios}
\frac{\Delta{}t_{QED}}{\Delta{}t_{CFL}}\sim\frac{10n}{a} \quad\quad \frac{\Delta{}t_{QED}}{\Delta{}t_D}\sim\frac{100}{a}\left(\frac{n_e}{n_c}\right)^{1/2}\left(\frac{m_ec^2}{k_bT_e}\right)^{1/2}
\end{equation}

Here $k_bT_e$ is the thermal energy of the electrons in the plasma.  In 
typical laser-solid simulations $\Delta{}t_D < \Delta{}t_{CFL}$, 
$n_e/n_c\sim O(10^3)$ and $m_ec^2/k_bT_e\sim O(10^3)$.  Therefore we require 
$a>O(10^5)$ 
for $\Delta{}t_{QED}$ to be the limiting constraint.  In laser-gas interactions 
it is usually the case that $\Delta{}t_{CFL} < \Delta{}t_{D}$.  In this case, 
if realtively coarse spatial resolution is used ($n=10$), one requires that 
$a>O(10^2)$ for $\Delta{}t_{QED}$ to set the time-step.

\section{Testing the Monte-Carlo Algorithm} 
\label{testing}

In this section the Monte-Carlo algorithm will be tested in two field 
configurations discussed in section \ref{master_equations}: a constant 
magnetic field and a circularly polarised electromagnetic wave.  We consider 
an electron bunch where the electrons initially all have energy 
$\gamma_0m_ec^2$, 
moving perpendicular to the electromagnetic fields (and counter-propagating 
relative to the wave in the second configuration).  The field strengths 
and values of $\gamma_0$ for each test case are given in table 
\ref{test_info}.  In each case we compare $\Phi_{-}$, $\Phi_{\gamma}$ \& 
$\Phi_{+}$ obtained from Monte-Carlo simulations to direct numerical 
solution\footnote{By first-order upwinding, ensuring the time-step is 
$<10^{-1}\Delta{}t_{QED}$.} of equations (\ref{master_electrons}) \& 
(\ref{master_photons}).  A comparison is also made 
to a `deterministic' emission model, where 
$\Phi_{-}(\gamma,t)=\delta[\gamma-\gamma_d(t)]$ and $\gamma_d(t)$ is the 
solution to the equation of motion (\ref{Losses_classical}) and to a classical 
model where $\gamma_d$ is calcualted assuming $g(\eta)=1$.

\begin{table}[b]
\centering
\begin{tabular}{c c c c c}
\hline\hline
Test & $\gamma_0$ & $|\mathbf{E}|/E_s$ & $c|\mathbf{B}|/E_s$ & $\eta_0$ \\ 
[0.5ex] 
\hline
$1$ & $1000$  & $0$  & $1\times10^{-3}$ & $1$ \\
$2$ & $4120$ & $1.22\times10^{-4}$  & $1.22\times10^{-4}$ & $1$ \\
$3$ & $1000$ & $0$ & $9\times10^{-3}$ & $9$ \\
\hline
\end{tabular}
\caption{\label{test_info} Details of each test case: in cases 1 \& 3 
electrons propagate perpendicular to a constant magnetic field, in case 2 
they counter-propagate relative to a circularly polarised plane 
electromagnetic wave.  $\eta_0$ is its initial value of $\eta$.}
\end{table}     

\begin{figure}
\centering
\includegraphics[scale=0.9]{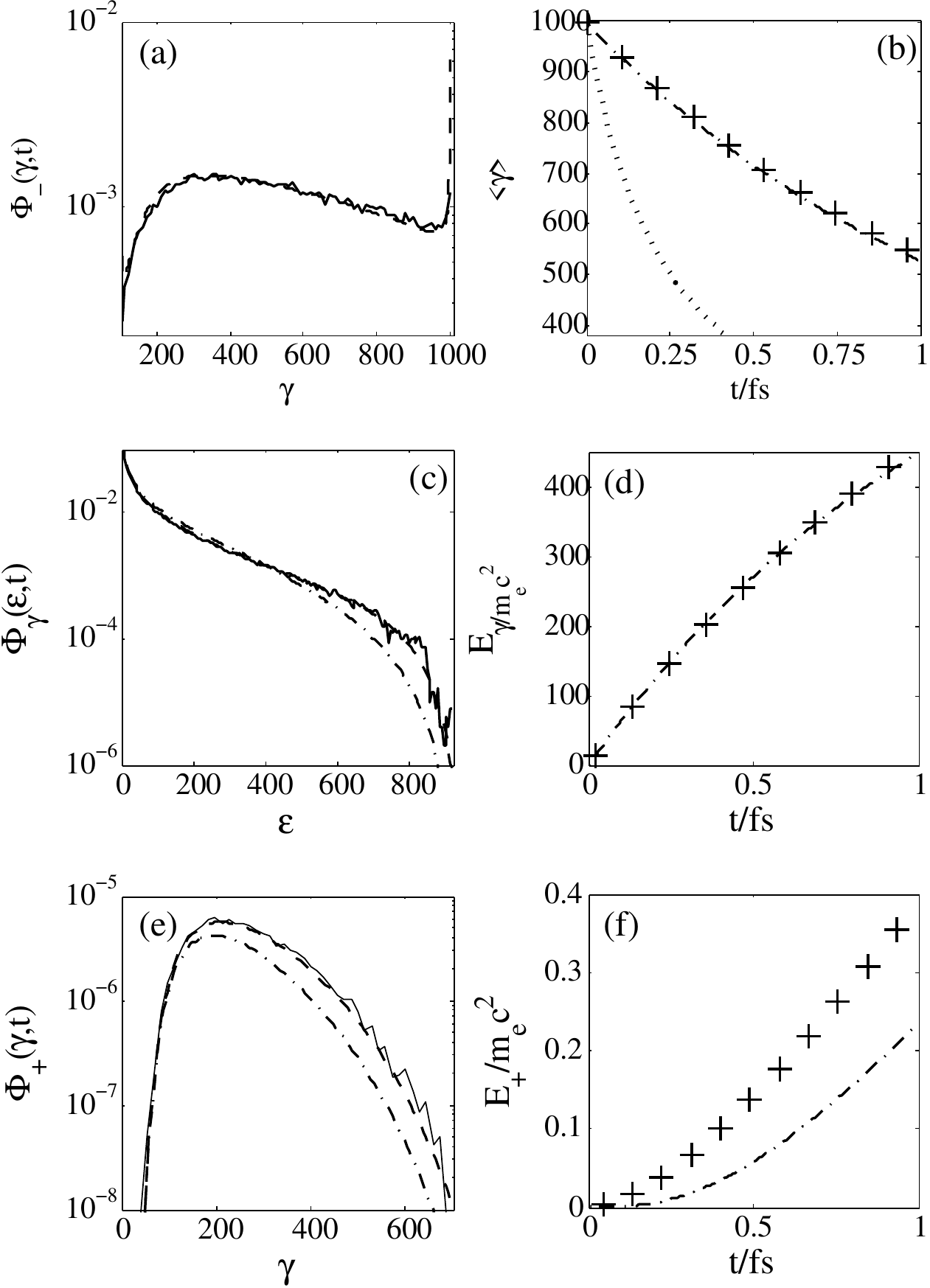}
\caption{\label{quantum_test_Bfield} Results for test problem 1: an electron 
bunch with $\gamma_0=1000$ moving perpendicular to a 
magnetic field of strength $B=10^{-3}E_s/c$.  (a) 
$\Phi_{-}(\gamma,t_0=1\mbox{fs})$ reconstructed from $10^5$ Monte-Carlo 
trajectories (solid line) compared to the result of direct numerical 
solution of equation (\ref{master_electrons}) 
(dashed line).  (b) $\gamma$ averaged over $10^5$ Monte-Carlo trajectories 
(crosses), the solution for deterministic losses (dot-dashed line) and the 
classical solution (dotted line). (c) 
$\Phi_{\gamma}(\gamma,t_0=1\mbox{fs})$ from the Monte-Carlo simulation 
(solid line) compared to the result of direct numerical 
solution of equation(\ref{master_photons}) (dashed line) 
and the spectrum produced by the deterministic model (dot-dashed line). 
(d) Total energy radiated as photons from the Monte-Carlo simulation (crosses) 
and for deterministic emission (dot-dashed line).  (e) \& (f)  Equivalent plots 
for positrons, where $10^7$ electrons were used to obtain the 
Monte-Carlo results.}
\end{figure}

\begin{figure}
\centering
\includegraphics[scale=0.9]{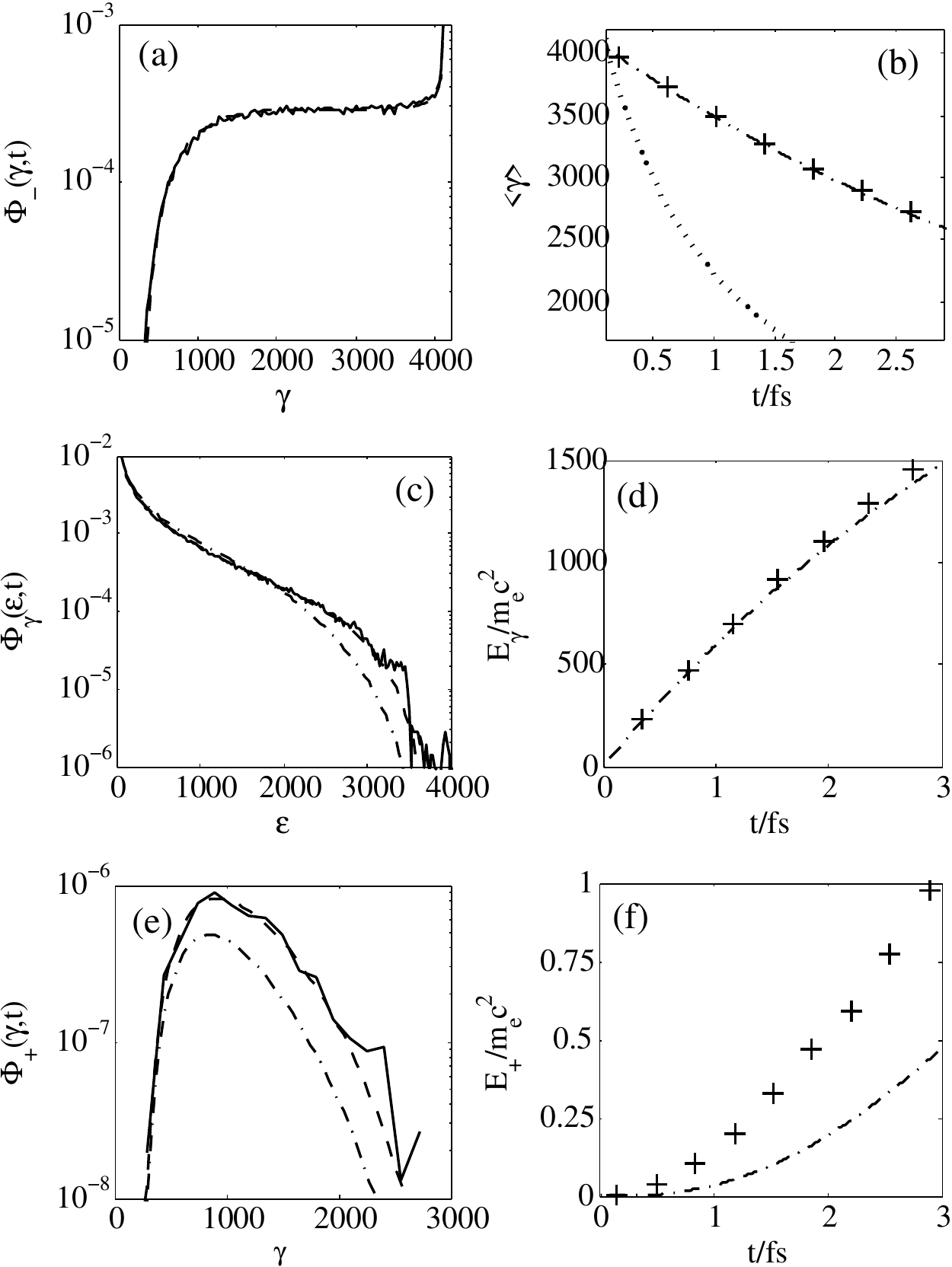}
\caption{\label{quantum_test_cplaser} Results for test problem 2: an  
electron bunch with $\gamma_0=4120$ counter-propagating relative
 to a circularly polarised electromagnetic wave with $a=50$.  (a) 
$\Phi_{-}(\gamma,t_0=3\mbox{fs})$ reconstructed from $10^5$ Monte-Carlo 
trajectories (solid line) compared to the result of direct numerical solution
 of equation (\ref{master_electrons}) (dashed line).  
(b) $\gamma$ averaged over $10^5$ Monte-Carlo trajectories (crosses), the 
solution for deterministic losses (dot-dashed line) and the classical solution 
(dotted line). (c) $\Phi_{\gamma}(\gamma,t_0=3\mbox{fs})$ from the Monte-Carlo 
simulation (solid line) compared to the result of direct numerical solution of 
equation (\ref{master_photons}) (dashed line) and the spectrum produced by the
deterministic model (dot-dashed line). (d) Total energy radiated as photons 
from the Monte-Carlo 
simulation (crosses) and the spectrum produced by the deterministic model 
(dot-dashed line).  (e) \& (f)  Equivalent plots for positrons, where $10^6$ 
electrons were used to obtain the Monte-Carlo results.}
\end{figure}

Figures \ref{quantum_test_Bfield} \& \ref{quantum_test_cplaser} show results 
for test problems 1 \& 2, where the initial $\eta$, $\eta_0=1$.  The results 
can be summarised as follows.  $\Phi_{-}$, $\Phi_{\gamma}$ \& $\Phi_{+}$ 
reconstructed from the Monte-Carlo code agree well with those obtained by 
direct numerical solution of equations (\ref{master_electrons}) \& 
(\ref{master_photons}).  This demonstrates that sufficient particles are 
used in the Monte-Carlo simulations to adequately sample the energy 
distributions.  $\Phi_{-}$ is extremely broad with 
$\sigma\sim\langle\gamma\rangle$ and so $\delta[\gamma-\gamma_d(t)]$ is an 
extremely poor fit to the distribution.  As a result the deterministic 
emission model 
fails to correctly predict the high energy tail in $\Phi_{\gamma}$.  The 
deterministic model does however correctly determine the total energy radiated 
as photons 
and the average electron trajectory.   The classical model predicts that the 
electrons radiate far too much energy.  Pair production is sensitive to the 
high energy tail tail in $\Phi_{\gamma}$ and so the deterministic model fails to 
predict the total energy emitted as positrons as well as the positron spectrum.
The Monte-Carlo algorithm produces the same positron spectrum as the
direct solution for $\Phi_{+}$ and so also the correct value for the total 
energy emitted as positrons. 

\begin{figure}
\centering
\includegraphics[scale=0.83]{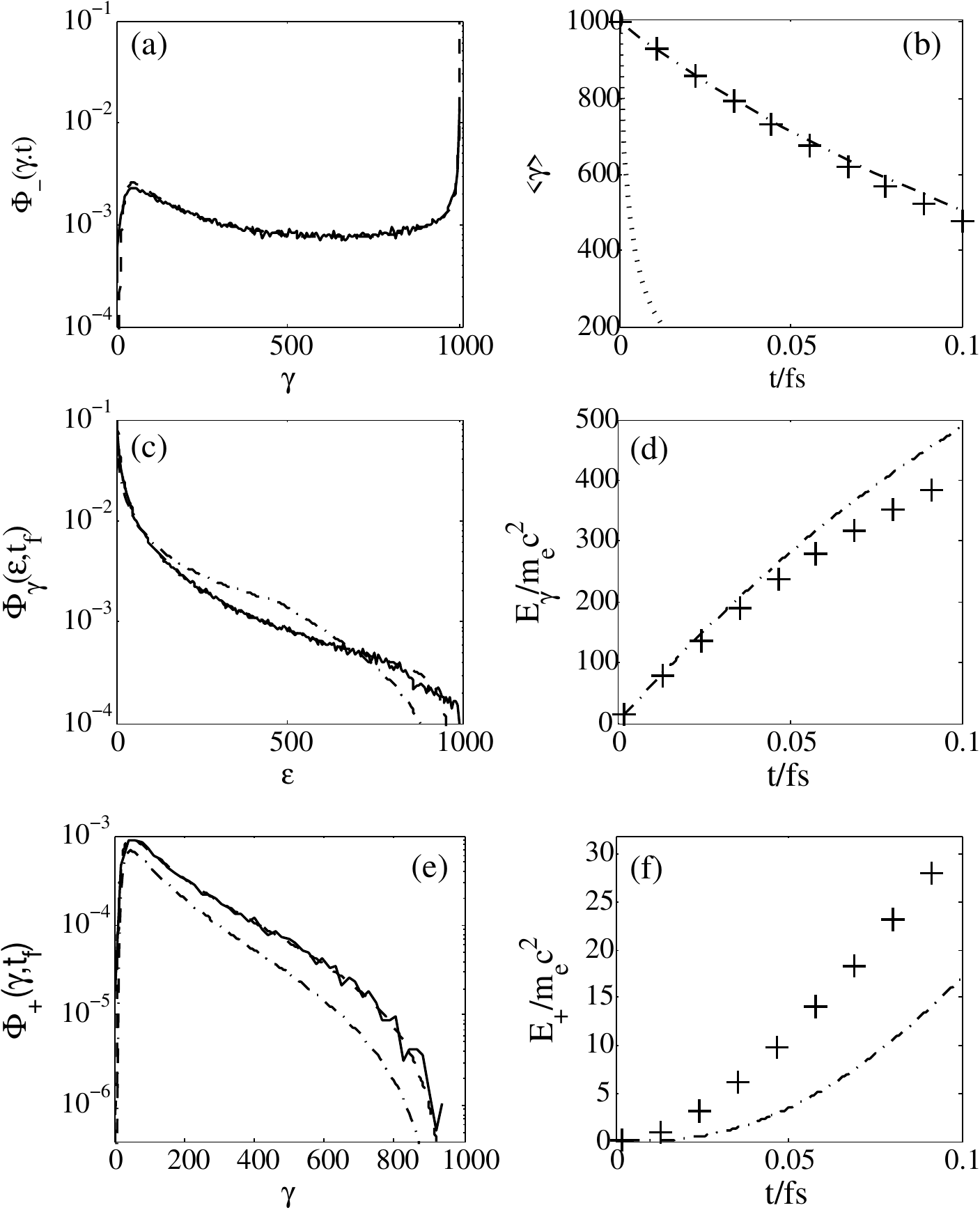}
\caption{\label{v_quantum_test_Bfield}  Results for test problem 3: an electron
 bunch with $\gamma_0=1000$ propagating perpendicular to a magnetic field of 
strength $B=9\times10^{-3}E_s/c$.  (a) $\Phi_{-}(\gamma,t_0=0.1\mbox{fs})$ 
reconstructed from $10^5$ Monte-Carlo trajectories (solid line) compared to 
the result of direct numerical solution of equation (\ref{master_electrons}) 
(dashed-line).  (b) $\gamma$ averaged over $10^5$ Monte-Carlo trajectories 
(crosses), the solution from the deterministic emission model (dot-dashed line) 
and the classical solution (dotted line). (c) 
$\Phi_{\gamma}(\gamma,t_0=0.1\mbox{fs})$ from the Monte-Carlo simulation 
(solid line) compared to the result of direct numerical solution of equation 
(\ref{master_photons}) (dashed line) and the spectrum produced by the 
deterministic emission model (dot-dashed line). (d) Total energy radiated as 
photons from the Monte-Carlo simulation (crosses) and the deterministic 
emission model (dot-dashed line).  (e) \& (f)  Equivalent plots for positrons.}
\end{figure}

Figure \ref{v_quantum_test_Bfield} shows the results for test problem 3.  In 
this case $\eta_0=9$.  Similarly to test problems 1 \& 2 the distributions 
$\Phi_{-}$, $\Phi_{\gamma}$ \& $\Phi_{+}$ obtained from the Monte-Carlo emission 
algorithm agree with those from direct solution of equations 
(\ref{master_electrons}) \& (\ref{master_photons}).  However, in this case the 
average electron energy and energy radiated as photons differ from the 
deterministic model.  This is because at such high $\eta$ pairs are no longer a 
minority species and so contribute to the average electron energy and 
to the radiation of gamma-rays.

\subsection{Accuracy, Numerical Convergence \& Energy Conservation}

\begin{figure}
\centering
\includegraphics[scale=0.9]{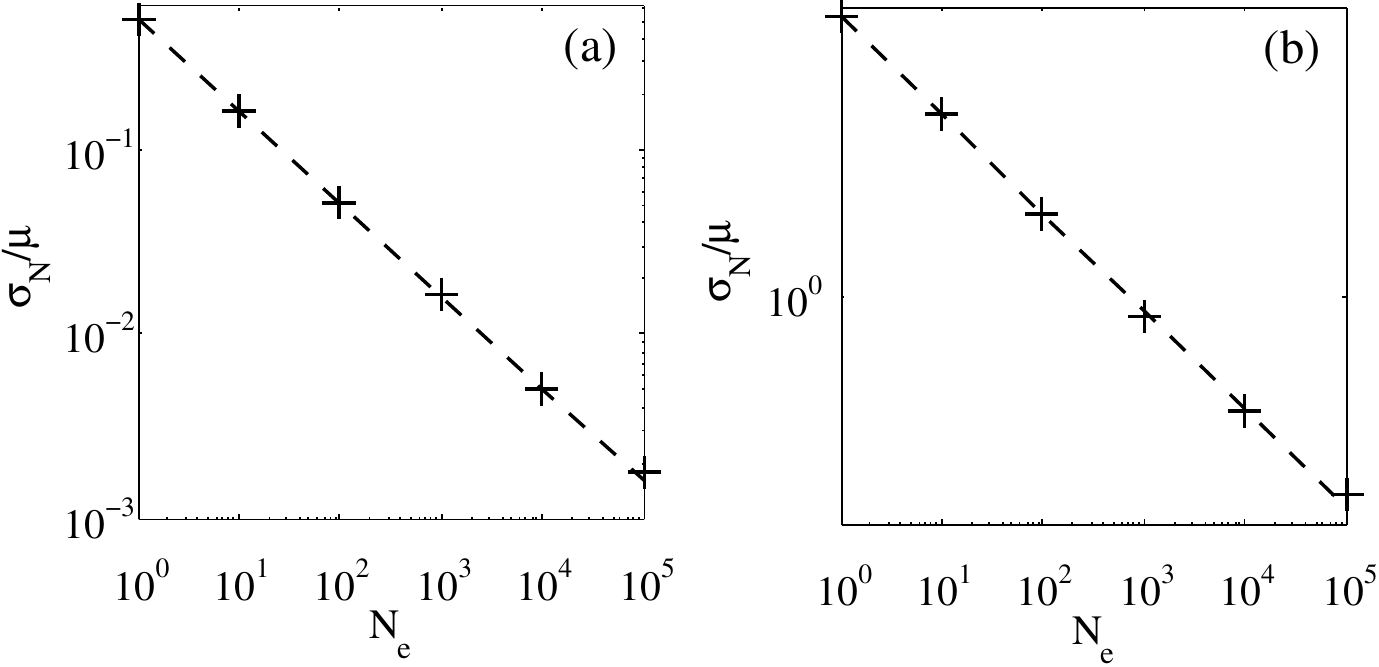}
\caption{\label{numerical_convergence} (a) Reduction in $\sigma_N/\mu$ for the 
total energy radiated as photons in test problem 1 with number of electrons 
initially present in the simulation $N_e$ (crosses) \& the $1/\sqrt{N_e}$ 
scaling (dashed line).  (b) The equivalent plot for positrons.}
\end{figure}

\begin{figure}
\centering
\includegraphics[scale=0.85]{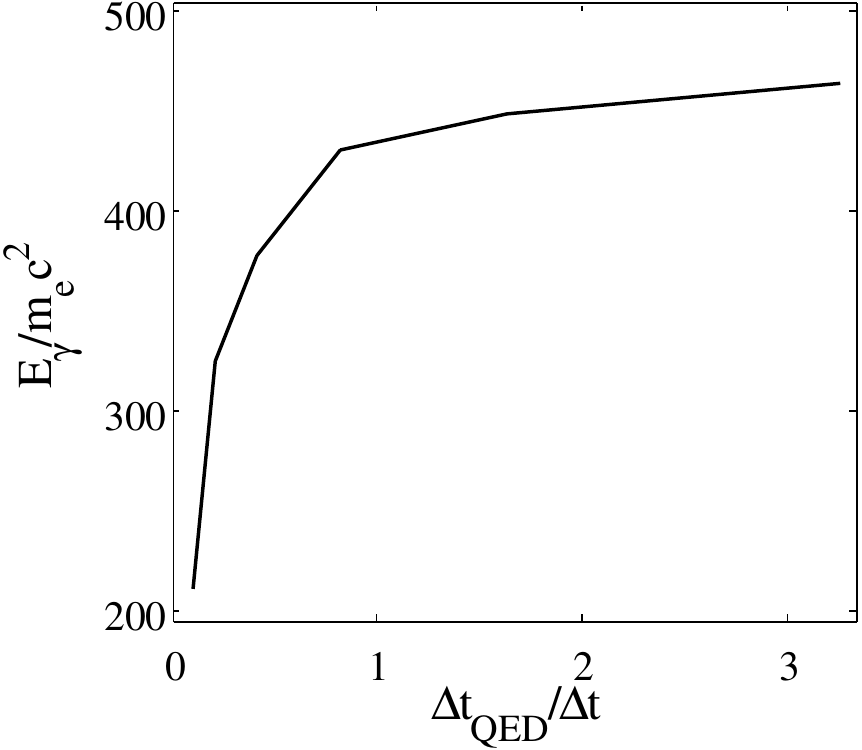}
\caption{\label{convergence_timestep} Convergence of energy emitted as gamma-ray photons $E_{\gamma}$ with decreasing time-step $\Delta{}t$ in test problem 1.}
\end{figure}

To investigate the convergence of the numerical solution, test problem 1 was 
repeated and the number of macroelectrons ($N_e$) and the time-step 
($\Delta{}t$) varied.  Figures \ref{numerical_convergence}(a) \& (b) show the 
decrease in the coefficient of variation $\sigma_N/E$ in the total energy 
radiated as gamma-ray photons and positrons with increasing $N_e$.  
$\sigma_N/E$ scales as $1/\sqrt{N_e}$.  Many more macroelectrons are required 
to resolve positron than photon production due to the lower emission rate.  
Another requirement for obtaining an accurate solution is that 
$\Delta{}t<\Delta{}t_{QED}$.  Figure \ref{convergence_timestep} shows how the 
energy 
radiated as photons per particle in the Monte-Carlo simulation converges as 
$\Delta{}t$ is varied between 
$\Delta{}t=10\Delta{}t_{QED}$ and $\Delta{}t=0.3\Delta{}t_{QED}$.  The error in 
$\epsilon_{\gamma}$ decreases quickly with decreasing time-step; when 
$\Delta{}t=0.6\Delta{}t_{QED}$ the solution has converged to a reasonable level 
of accuracy (3\%).  

In section \ref{numerical_constraints} we demonstrated that the numerical 
scheme does not conserve energy.  Table \ref{delta_energy_1} shows the energy, 
divided by the initial number of electrons in the bunch, in electrons, 
photons and positrons at the beginning 
and end of simulations in the constant B-field test cases (1 \& 3), where the 
electrons \& positrons gain no energy from the classical fields.  The last 
column shows the error in energy conservation summed over all particles in
the simulation divided by $N_e$.  This shows that energy is conserved to a 
high degree of 
accuracy ($<0.01$\%) in each simulation.

\begin{table}
\centering 
\begin{tabular}{c c c c c}
\hline\hline
Test & Component & Energy/$m_ec^2$ ($t=0$) & Energy/$m_ec^2$ ($t=t_0$) & 
$\Delta$(Energy/$m_ec^2$) \\ [0.5ex] 
\hline
1 &Electron & 1000.0  & 544.8 & -455.2 \\
 & Photon & 0 & 454.8  & 454.8 \\
 & Positron & 0 & 0.4 & 0.4 \\
 & Total & 1000.0 & 1000.0 & 0.001 \\
\hline
3 & Electron & 1000.0  & 562.1 & -437.9 \\
 & Photon & 0 & 406.3  & 406.3 \\
 & Positron & 0 & 31.7 & 31.7 \\
 & Total & 1000.0 & 1000.0 & 0.006 \\ 
\hline
\end{tabular}
\caption{Energy in each component in Monte-Carlo simulations of test problems 
1 (with $N_e=10^7$) \& 3 (with $N_e=10^6$).  The `Total' in $\Delta$(Energy) 
refers to the total error in energy conservation summed over all particles in 
the simulation.}
\label{delta_energy_1}
\end{table}

\section{Discussion}

In section \ref{emission_section} a quasi-classical model for the QED emission 
processes was described; in particular kinetic equations describing the 
evolution of distribution functions were derived for 
electrons, positrons \& photons.  It was shown that there is a close 
correspondence between the moments of these equations and the equation 
of motion using a deterministic model for the radiation reaction force.  
Such 
a deterministic emission model has been adopted by several authors 
\cite{Kirk_09,Sokolov_10,Zhidkov_02}.  When the probabilistic QED model and the 
deterministic 
model were compared in section \ref{testing} it was found that the 
deterministic model did not correctly predict the emitted photon or positron 
spectra.  Although it was found that in the particular cases considered here 
the the deterministic model did correctly predict the total energy radiated by 
the electrons when pair production was negligible\footnote{For 
$\eta\sim1$, $\mathcal{P}(\eta)$ goes approximately as $\sim\eta$ and so 
$d\langle{}\gamma\rangle/dt \approx d\gamma_c/dt$, despite the fact that 
for $\eta\sim1$ $\Phi_{-}$ is a broad distribution.}.  In cases more relevant 
to experiments, such as a laser pulse striking a solid target 
\cite{Ridgers_12,Duclous_11} or an electron beam interacting with a laser 
pulse with a 
Gaussian temporal envelope, calculations suggest larger differences between 
the probabilistic \& deterministic models. This suggests that the probabilistic
 Monte-Carlo emission algorithm described here is preferable to a deterministic
 emission model. 

The Monte-Carlo scheme introduces additional numerical constraints on the 
time-step and the minimum number of macroparticles that can be used in 
QED-PIC simulations.  The time-step must be smaller than $\Delta{}t_{QED}$.  
In section \ref{numerical_constraints} it was shown that this only limits the 
time-step in a QED-PIC simulation of low density plasmas when $a>10^2$.  The 
total energy emitted as 
photons \& pairs converges with number of macroelectrons in the simulation 
($N_e$) as $\sigma_N/E=(1/\sqrt{N_e})\sigma/E$.  The number of macroelectrons 
required for $\sigma_N/E$ to converge to an acceptable level depends on 
$\sigma/E$ which depends on the rate of emission.  If the rate of emission is 
reduced $\sigma/E$ is increased.  This explains why it was found that many 
more macroparticles were required to get a reasonable degree of convergence 
in positron emission than in photon emission for $\eta_0=1$, as the rate of 
positron production is considerable lower than that for photon production at 
this $\eta$.  
However, this is precisely the case when pair production does not affect the 
plasma dynamics.  For $\eta_0=9$, as examined in figure 
\ref{v_quantum_test_Bfield}, the rate of pair production is higher, pair 
production does affect the plasma dynamics and in this case the same number 
of macroparticles is required to resolve positron emission as photon emission.

The emission model breaks down when the fields can no longer be considered as 
quasi-static or when the laser's electric field becomes close to the Schwinger 
field.  The former condition is typically only satisfied in 
situations where emission is unimportant and the latter requires laser pulses 
of extremely high intensity ($10^{28}$Wcm$^{-2}$) unlikely 
to be reached in the near term.  We therefore conclude that the emission model 
outlined here is applicable to a very wide range of laser intensities.  

Finally, we note that processes not included in the model might be important 
under special conditions. For example, when the rate of pair production is 
very small, it is dominated by the second-order (in $\alpha_{f}$) trident 
process. A comprehensive discussion of higher order processes can be found 
in Ref. \cite{DiPiazza_12}. Also, in addition to the Coulomb collisions between 
electrons and ions in the plasma usually included in PIC simulations, 
additional collisional processes could play role.  For example one could 
include collisions between: gamma-ray photons and electrons/positrons (Compton 
scattering); electrons and positrons (annihilation); 
electrons/positrons and ions/atoms (bremsstrahlung or Trident pair production 
in the electric fields of the nuclei); gamma-ray photons and atoms 
(Bethe-Heitler pair production).  A full investigation of the relative 
importance of these effects is beyond the scope of this paper. 

\section{Conclusions}

When laser pulses of intensity $>10^{21}$Wcm$^{-2}$ interact with 
ultra-relativistic electrons a significant amount of the electron's energy is 
converted to gamma-ray photons \& pairs.  We have shown that a probabilistic 
Monte-Carlo 
algorithm best simulates the emission and that such an algorithm can be 
coupled to a PIC code to simulate QED-plasmas.  By contrast 
a deterministic treatment of the emission processes only correctly describes 
the evolution 
of the particle spectra when pair production can be neglected and so is only 
valid over a relatively narrow range of laser intensities.  
We therefore conclude that QED-PIC codes, using the Monte-Carlo emission 
algorithm described here, will provide a valuable tool for simulating high 
intensity laser-plasma interactions at today's highest intensities and beyond.

\ \\

\textbf{Acknowledgements}
\ \\

This work was funded by the UK Engineering and Physical Sciences Research 
Council (EP/G055165/1 \& EP/G054940/1).  We would like to thank Brian Reville 
\& Alexander Thomas for many useful discussions.

\appendix
\section{Classical \& Quantum Synchtrotron Emissivity}
\label{photon_emissivity_section}

The quantum synchrotron function is given by Sokolov and Ternov 
\cite{Sokolov_68} eq. (6.5). In our notation it is, for $\chi<\eta/2$
\begin{equation}
F(\eta,\chi)=\frac{4\chi^2}{\eta^2}y K_{2/3}(y)+
\left(1-\frac{2\chi}{\eta}\right)y\int_y^\infty\,\textrm{d}t\,K_{5/3}(t)
\end{equation}

where $y=4\chi/[3\eta(\eta-2\chi)]$ \& $K_n$ are modified Bessel functions 
of the second kind.  For $\chi\ge \eta/2$, $F(\eta,\chi)=0$.

\begin{figure}
\centering
\includegraphics[scale=1.05]{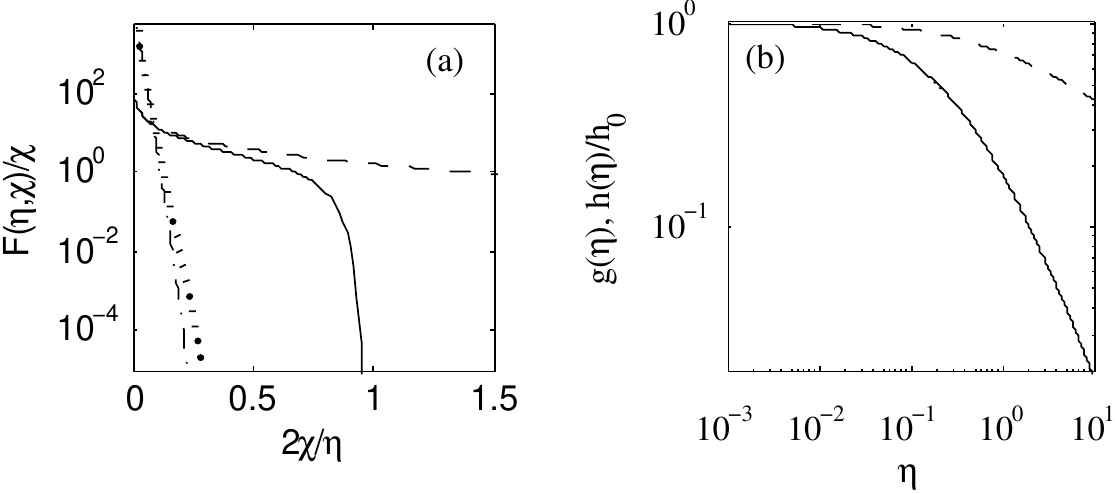}
\caption{\label{h_g_fig}(a) $F(\eta,\chi)/\chi$ and the equivalent classical
spectrum plotted for $\eta=0.01$ (dash-dot and dotted lines respectively) \& 
$\eta=1$ (solid and dashed lines respectively).  (b) $g(\eta)$ (solid line) \&
$h(\eta)$ (dashed line).}
\end{figure}

In the classical limit $\hbar\rightarrow0$ the quantum synchrotron spectrum 
reduces to the classical synchrotron spectrum  
$F(\eta,\chi)\rightarrow y_c\int_{y_c}^{\infty}du K_{5/3}(u)$; $y_c=4\chi/3\eta^2$. 
 For comparison the classical and quantum synchrotron spectra are plotted for
$\eta=0.01$ \& $\eta=1$ in figure \ref{h_g_fig}(a).  The classical spectrum
extends beyond the maximum possible photon energy, set by $2\chi/\eta=1$.

As stated in section \ref{radiation_section}, the modification to the spectrum 
leads to a reduction in the radiated power by a factor $g(\eta)$, where a fit 
to this function was given.  The photon emissivity is also reduced by a factor 
of $h(\eta)/h_0$.  $g(\eta)$ \& $h(\eta)$ are expressed in terms of 
$F(\eta,\chi)$ as

\begin{equation}
g(\eta)=\frac{3\sqrt{3}}{2\pi\eta^2}\int_0^{\eta/2}d\chi F(\eta,\chi) \quad\quad
h(\eta) = \int_0^{\eta/2}d \chi \frac{F(\eta,\chi)}{\chi}
\end{equation}

These functions are plotted in figure \ref{h_g_fig}(b).  Here 
$h(\eta)$ has been normalised to the classical value 
$h_0=5.24$.  Quantum corrections to the photon emission become important when 
$g(\eta)$ and $h(\eta)/h_0$ deviate from unity.  

\section{Pair Emissivity}
\label{pair_emissivity_section}

The approximate form of the function controlling the rate of pair production 
used here is \cite{Erber_66}

\begin{figure}
\centering
\includegraphics[scale=1.1]{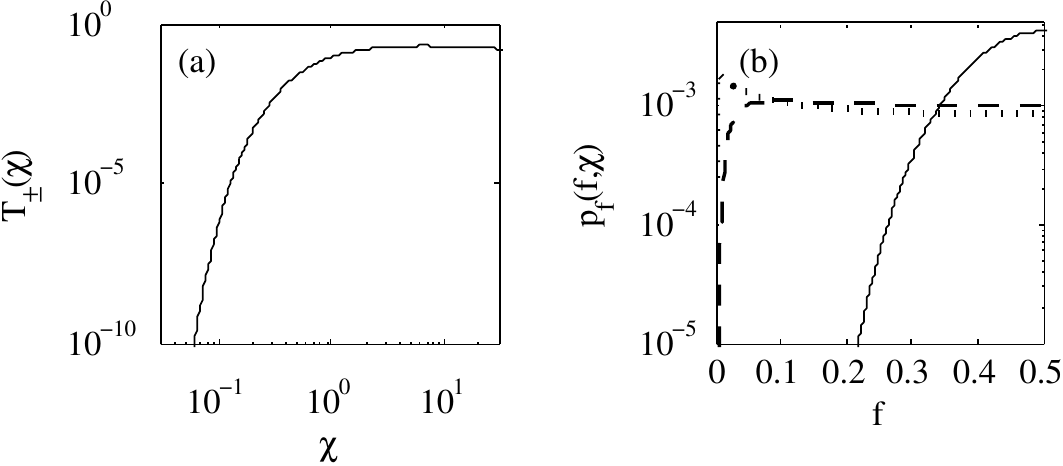}
\caption{\label{T_pm_fig}. (a) $T_{\pm}(\chi)$.  (b) $p_f(f,\chi)$ plotted for
$\chi=0.1$ (solid line) $\chi=10$ (dashed line) \& $\chi=100$ (dotted line).}
\end{figure}

\begin{equation}
T_{\pm}\approx 0.16\frac{K_{1/3}^2[2/(3\chi)]}{\chi}
\end{equation}

This function is plotted in figure \ref{T_pm_fig}(a).  Note the extremely 
rapid increase with $\chi$; for low $\chi$ 
$T_{\pm}(\chi)\propto \exp[-2/(3\chi)]$.  For high 
$\chi$ $T_{\pm}(\chi)$ falls off as $\chi^{-1/3}$.

The function controlling the distribution of the photon energy between the 
electron and positron in the pair, $p_f(f,\chi)$, is given by 
\cite{Daugherty_83}

\begin{align}
p_f(f,\chi) = \frac{2+f(1-f)}{f(1-f)}K_{2/3}\left[\frac{1}{3\chi f(1-f)}\right]\frac{1}{k(\chi)}
\end{align}

Where $k(\chi)$ is a normalisation constant such that 
$\int_0^1 df p_f(f,\chi)=1$. Note that $p_f(f,\chi)=p_f(1-f,\chi)$ and so is 
symmetrical in $f$ about $f=0.5$.

In the limits $\chi \ll 1$ \& $\chi \gg 1$ $p_f$ approaches

\begin{align}
p_f(f,\chi)\approx \frac{2+f(1-f)}{(\chi f(1-f))^{1/2}}\exp\left[ -\frac{1}{3\chi f(1-f)}\right]\frac{1}{k(\chi)} \quad\quad \chi\ll1 \\
p_f(f,\chi)\approx\frac{2+f(1-f)}{(\chi f(1-f))^{1/3}}\frac{1}{k(\chi)} \quad\quad \chi\gg \frac{1}{f(1-f)}
\end{align}

Therefore, $p_f(f,\chi)$ is sharply peaked at $f=0.5$ for $\chi\ll1$ and peaked
at $f \approx 0$ \& $f \approx 1$ for $\chi\gg1$.  This is demonstrated by 
figure \ref{T_pm_fig}(b), where $p_f(f,\chi)$ is plotted for $\chi=0.1,1$ \& 
100.

\end{document}